\documentclass{emulateapj}
\pdfoutput=1
\usepackage{epsf}
\usepackage{epsfig}
\DeclareGraphicsExtensions{.jpg,.pdf,.png,.eps,.ps}

\usepackage{amsmath}
\usepackage{natbib}
\usepackage{graphicx}
\usepackage{color}
\usepackage{threeparttable}
\usepackage{xspace}
 \newcommand{\LCDM}{\mbox{$\Lambda$CDM}\xspace}

 \newcommand{\ltsima}{$\; \buildrel < \over \sim \;$}
 \newcommand{\ltsim}{\lower.5ex\hbox{\ltsima}}

 \newcommand{\msun}{M_{\odot}}
 \newcommand{\sqdeg}{\ensuremath{\mathrm{deg}^2}}
 
 \newcommand{\eg}{\textit{e.g.}}

 \newcommand{\des}{DES-\textit{base}}
 \newcommand{\desagg}{DES-\textit{agg}}
 \newcommand{\sptsz}{\rm{SPT-SZ}}
 \newcommand{\sptg}{\rm{SPT-3G}}
 \newcommand{\microkam}{$\mu$K-arcminute}
 \newcommand{\zavg}{$z_{\rm{avg}}$}
 \newcommand{\dcom}{$r_{\rm{com}}$}

\newcommand{\<}{\langle}
\renewcommand{\>}{\rangle}
\newcommand{\be}{\begin{equation}}
\newcommand{\ee}{\end{equation}}
       
\def\ba#1\ea{\begin{align}#1\end{align}}

\begin{document}

\title{
Prospects for measuring the relative velocities of galaxy clusters in photometric surveys using the kinetic Sunyaev-Zel'dovich Effect}
\author{Ryan Keisler\altaffilmark{1,2,3}, Fabian Schmidt\altaffilmark{4,5}}

\altaffiltext{1}{Kavli Institute for Cosmological Physics,
University of Chicago, 5640 South Ellis Avenue, Chicago, IL~~60637, USA}
\altaffiltext{2}{Department of Physics,
University of Chicago,
5640 South Ellis Avenue, Chicago, IL~~60637, USA}
\altaffiltext{3}{Hubble Fellow}
\altaffiltext{4}{Department of Astrophysical Sciences, Princeton University,
Princeton, NJ~~08540, USA}
\altaffiltext{5}{Einstein Fellow}

\email{rkeisler@uchicago.edu, fabians@astro.princeton.edu}

\begin{abstract}
We consider the prospects for measuring the pairwise kinetic Sunyaev-Zel'dovich (kSZ) signal from galaxy clusters discovered in large photometric surveys such as the Dark Energy Survey (DES).  We project that the DES cluster sample will, in conjunction with existing mm-wave data from the South Pole Telescope (SPT), yield a detection of the pairwise kSZ signal at the 8-13$\sigma$ level, with sensitivity peaking for clusters separated by $\sim$100 Mpc distances.  A next-generation version of SPT would allow for a 18-30$\sigma$ detection and would be limited by variance from the kSZ signal itself and residual thermal Sunyaev-Zel'dovich (tSZ) signal.  Throughout our analysis we assume photometric redshift errors, which wash out the signal for clusters separated by $\lesssim$50 Mpc; a spectroscopic survey of the DES sample would recover this signal and allow for a 26-43$\sigma$ detection, and would again be limited by kSZ/tSZ variance.  Assuming a standard model of structure formation, these high-precision measurements of the pairwise kSZ signal will yield detailed information on the gas content of the galaxy clusters.  Alternatively, if the gas can be sufficiently characterized by other means (\eg~using tSZ, X-ray, or weak lensing), then the relative velocities of the galaxy clusters can be isolated, thereby providing a precision measurement of gravity on 100 Mpc scales.  We briefly consider the utility of these measurements for constraining theories of modified gravity.


\end{abstract}

\section{Introduction}
\label{sec:introduction}

Measurements of the peculiar velocities of massive halos, in conjunction with expansion history measurements, can be used to constrain cosmological models and test gravity on cosmological scales \citep{kosowsky09, reyesetal10}.  The kinetic Sunyaev-Zel'dovich (kSZ) effect \citep{sunyaev72} effect provides a means to measure peculiar velocities.  CMB photons gain or lose energy after Thomson scattering on free electrons moving with respect to the CMB, and the resulting shift in the CMB temperature in the direction of the electrons is known as the kSZ effect.

The kSZ effect was recently measured at the 3.8$\sigma$ level \citep{hand12} using mm-wave data from the Atacama Cosmology Telescope (ACT) in conjunction with the BOSS spectroscopic catalog \citep{sdss12}.  This was a measurement of the \textit{pairwise} kSZ signal; two massive halos will tend to fall towards each other, and the resulting CMB temperature difference is the pairwise kSZ signal.

The cosmological utility of a catalog of kSZ-derived velocities was studied in \cite{bhattacharya07} and \cite{kosowsky09}, and the prospects for measuring growth using the angular correlation of galaxies found in the Dark Energy Survey\footnote{http://www.darkenergysurvey.org} (DES) was studied by \citet{ross11}.  In this letter we consider the prospects for detecting the pairwise kSZ signal using existing or future mm-wave data from the South Pole Telescope (SPT, \citet{carlstrom11}) and galaxy clusters discovered by the DES, which is coming online at the time of writing.  We simulate photometric redshift errors and realistic mm-wave data, including the noise introduced by the SZ effects, and find that these datasets will yield strong detections of the pairwise kSZ signal.

\section{Analysis}
\label{sec:analysis}

Our analysis relies on the simulations presented in \cite{sehgal10}, hereafter S10.  This work constructed simulated maps of the mm-wave sky with contributions from the CMB, the thermal and kinetic SZ effects, and emissive sources.  The pieces of S10 that are relevant for our analysis are the halo catalog and the simulated maps of the kSZ and tSZ signals.  The SZ signals were calculated per cluster in post-processing; the gravitational potential was calculated according to the N-body particle density, and the gas density and pressure were calculated assuming hydrostatic equilibrium and a polytropic equation of state.  S10 calculated the kSZ signal after assigning a single, mass-averaged velocity to all gas associated with each cluster.
We scale the S10 tSZ map by a factor of 0.7, which is in broad agreement with the signals measured in recent tSZ cluster surveys \citep{benson11} and measurements of the tSZ power spectrum \citep{reichardt12b}.

\subsection{Cluster Sample}
\label{sec:clusters}

We construct two DES-like cluster samples from the S10 halo catalog.  The first, dubbed \des, includes clusters with $1\times10^{14}~\msun < M_{500c} < 3\times10^{14}~\msun$ and $0.2<z<1.0$.  The lower mass threshold, $M_{500c} = 1\times10^{14}~\msun$, corresponds to  $\sim$20 galaxies \citep{rykoff12}, which is the expected selection threshold for DES clusters.  We ignore scatter in mass (at fixed optical richness) and offsets between the optically-derived cluster center and the center of gas mass, both of which could potentially be serious problems for this analysis and would need to be modeled in a more realistic analysis.  The second, more aggressive sample, dubbed \desagg, has the same redshift range but a lower mass range: $5\times10^{13}~\msun < M_{500c} < 1.5\times10^{14}~\msun$.  Such low mass systems might be obtained in exchange for lower purity and higher mass scatter.

We have limited the upper mass range of these samples to create more uniform samples and to mitigate strong tSZ signals.  This requirement reduces the number density of clusters by $\sim$10\% per sample, leading to $\sim$4 (16) clusters per \sqdeg~in the \des~(\desagg) sample.

We assign photometric redshift errors to the cluster redshifts
according to $\sigma_z = 0.01\times\exp(z/2.5)$, approximating the photometric errors presented in \textit{The Supplements for the Dark Energy Survey White Paper}\footnote{https://des.fnal.gov/survey\_documents/DES-DETF/Supplements\_DES-DETF\_v1.6.pdf}.  The mean redshifts are the redshifts provided in the S10 catalog, which do not include distortions due to velocities.  Catastrophic redshift errors, which we have ignored, would dilute the average pairwise kSZ signal by roughly the rate of their occurrence, $\sim$1\% \citep{sun09c, bernstein10}, an insignificant amount.

\subsection{mm-wave Data}
\label{sec:mm}

We simulate observations of the mm-wave sky by the SPT.  We consider two surveys:

\begin{itemize}

\item{\sptsz~- the initial, 2500 \sqdeg~survey, for which we assume white noise levels of [40, 18, 80] \microkam~and beam widths of [1.6, 1.05, 0.85] arcminute FWHM at [95, 150, 220] GHz, and}

\item{\sptg~- a future, third-generation survey, which covers the same 2500 \sqdeg~area, and for which we assume white noise levels of  [4.3, 2.5, 4.3] \microkam~and beam widths of [1.6, 1.05, 0.85] arcminute FWHM at [95, 150, 220] GHz.}

\end{itemize}

While each of these surveys has or would have coverage at 95, 150, and 220 GHz, we use only the 150 and 220 GHz data in the analysis presented here (with one exception, discussed in Section \ref{sec:results}).  We assume a background of randomly-distributed, emissive sources with $\sqrt{C^{\rm{}}_\ell}=7$ (21.5) \microkam~at 150 (220) GHz that is perfectly spatially correlated between 150 and 220 GHz, consistent with the results of \cite{reichardt12b}.  We optimally combine the 150 and 220 GHz data as a function of multipole to minimize the total power due to instrumental noise and emissive sources, and the resulting noise power is defined as $N_\ell$.


We simulate mm-wave data by combining the SZ skies from the S10 simulations with Gaussian realizations of $N_\ell$, and Gaussian realizations of $C_\ell^{\rm{CMB}}$, the best-fit \LCDM CMB power spectrum as constrained by SPT+WMAP in \cite{keisler11}.  We use $n_{\rm{side}}=2^{13}$ Healpix\footnote{http://healpix.jpl.nasa.gov} maps.

\vskip 30pt
\subsection{Signal Estimation}
\label{sec:signal}

Next we estimate the pairwise kSZ signal of the DES-like cluster samples in the simulated mm-wave data.  The pairwise velocity is a function of cosmology, cluster separation, and redshift (see Section~\ref{sec:gravity}), and we thus wish to estimate the amplitude of the pairwise kSZ signal in bins of cluster separation and redshift.  This amounts to estimating in data $d$ the amplitude $T_0$ of a signal with spatial template $\mathcal{K}$ in the presence of stationary noise with power spectrum $M_\ell$.  \cite{haehnelt96} show that the optimal estimate of this amplitude is

\begin{equation}
\label{eq:est}
\hat{T}_0 = \frac{\sum_{\ell,m} \operatorname{Re}[d_{\ell m}\mathcal{K}^{*}_{\ell m}]M^{-1}_\ell}{\sum_{\ell,m} \operatorname{Re}[\mathcal{K}_{\ell m}\mathcal{K}^{*}_{\ell m}]M^{-1}_\ell}, 
\end{equation}
where $d_{\ell m}$ ($\mathcal{K}_{\ell m}$) is the harmonic transform of the data (signal).  $\hat{T_0}$ estimates the average pairwise ionized momentum of the cluster sample in a given bin of cluster separation and redshift.  To construct the signal template $\mathcal{K}$ for each bin, we assume that a cluster $j$ contributes to the observed kSZ signal from cluster $i$ with an amount given by 

\begin{equation}
T_{ij}(\theta) = T_{0}c_{ij}\beta(\theta) \equiv T_{0}c_{ij}\left(1+\frac{\theta^2}{\theta^2_c}\right)^{-1}
\end{equation}
where $\theta$ is the angle from cluster $i$ and $\theta_c$ is the angle subtended by $r_c$=0.36 Mpc (physical distance); we have assumed a $\beta$-model profile for the Thomson optical depth with $\beta=1$ and $r_c$=0.36 Mpc, which approximates the profiles of the DES-like clusters in S10.  The $c_{ij}$ factor, defined as $c_{ij}\equiv \mathbf{\hat{r}}_{ij} \cdot \frac{\mathbf{\hat{r}}_{i} + \mathbf{\hat{r}}_{j}}{2}$, accounts for the on-sky projection of the clusters' separation vector, where $\mathbf{r}_{i}$ is the comoving distance to cluster $i$, and $\mathbf{r}_{ij} \equiv \mathbf{r}_i - \mathbf{r}_j$.

We construct the kSZ signal template $\mathcal{K}$ per bin of redshift and cluster separation by populating a blank Healpix map with $\sum_{j} c_{ij}$ at the location $\mbox{\boldmath$\theta$}_i$ of each cluster $i$, for all pairs of clusters \{$ij$\} in this bin, and convolving this map by the $\beta$-model profile:

\be
\mathcal{K}(\mbox{\boldmath$\theta$}) = \left(\sum_{i,j}c_{ij}\delta(\mbox{\boldmath$\theta$}-\mbox{\boldmath$\theta$}_i)\right) \ast \beta(\theta).
\ee
Finally, we estimate the amplitude $\hat{T}_0$ of this template in the data $d$ according to Equation~\ref{eq:est}.  We repeat this procedure for several bins in pair-averaged redshift \zavg$\in$[\{0.2, 0.4\}, \{0.4, 0.7\}, \{0.7, 1.0\}], and cluster pair separation, \dcom$\in$[\{10, 30\}, \{30, 70\}, \{70, 130\}, \{130, 200\}, \{200, 300\}] Mpc.  The kSZ signal in the bin of largest separation may be affected by the 1000 comoving Mpc/$h$ size of the simulation box used in S10.

Many of the assumptions used to construct the signal template $\mathcal{K}$ (\eg~that all clusters within a bin of redshift and cluster separation have an identical gas profile) are inaccurate at some level, but these assumptions can be regarded as \textit{definitions} of the analysis technique.  The technique is then calibrated by duplicating the analysis on simulated skies, as was done here.  
The challenge, of course, will be modeling and marginalizing over the inadequacies of the simulated skies.

Another technique, adopted by \cite{hand12}, is to convolve the mm-wave map with an optimal spatial filter, and, for all cluster pairs $\{ij\}$ within some range of pair separation, measure 

\be
\hat{T}_0 = -\frac{\sum_{i<j}(T(\mbox{\boldmath$\theta$}_i) - T(\mbox{\boldmath$\theta$}_j))c_{ij}}{\sum_{i<j}c^2_{ij}}
\ee
where $T(\mbox{\boldmath$\theta$}_i)$ is the filtered temperature at the location of cluster $i$.  We have confirmed analytically that this is equivalent to the harmonic-space technique.

\subsection{Noise Estimation}
\label{sec:noise}

The S10 SZ maps are duplicated per octant of sky, and we assign one 2500 \sqdeg~SPT survey area, including unique photometric redshift errors, per octant.  The CMB, instrumental noise, and point source background are generated as Gaussian realizations of $M_\ell$.  The kSZ amplitude $T_0$ is estimated once per sphere, and the scatter per survey is $\sqrt{8}$ larger than the scatter per sphere.

We simulate $\ge$120 spheres per data scenario.  This ensemble of $T_0$ estimates is used to construct a covariance matrix $C$ in bins of cluster separation.  $C$ is calculated per redshift bin, per data scenario.

The noise introduced by photometric redshift errors, instrumental noise, the background of point sources, and the CMB
are accounted for naturally in this scheme, but there is additional noise introduced by the kSZ and tSZ signals.  Both of these arise from the finite number of clusters in the sample.  We emphasize that the tSZ signal is not a source of bias; the pairwise kSZ signal is differential, and the tSZ will average to zero.

Ideally the kSZ and tSZ noise would be estimated in the same manner as the other sources of noise --- per octant, per Healpix realization --- but we have only one independent octant of S10 SZ maps.  Instead, we estimate the kSZ and tSZ noise using a bootstrap technique.  We divide the cluster sample into eight redshift bins and measure the scatter in $T_0$ (at fixed cluster separation) across these redshift bins.  We translate this result to our main analysis by accounting for the different redshift bin widths and by assuming that the kSZ and tSZ covariance matrices have the same dimensionless bin-bin correlation structure as the main covariance matrix $C$ described above.  

One source of noise that is absent in our analysis is velocity substructure within a cluster.  
S10 assigned a single peculiar velocity to all gas associated with a particular cluster when calculating the kSZ map.  However, several works (\eg~\citet{nagai03}, \citet{holder04}, \citet{dolag13}) have shown that velocity substructure limits the accuracy with which mm-wave data can estimate a cluster's velocity.  The systematic floor is $\sim$50-100 km/s per cluster, or $\delta T$$\sim$2$~\mu$K for the clusters considered here.  This is subdominant for \sptsz~noise levels, and unlikely to change our results for \sptg~by more than 10\%.

We have also neglected uncertainty in $\sigma_z$, the width of the photometric error distribution.  The effect of photometric errors on the kSZ signal is a strong function of cluster pair separation and most strongly affects $\lesssim$50 Mpc separations, as shown in the bottom panels of Figure~\ref{fig:snr}.  For this reason we expect the marginalization over $\sigma_z$ to only mildly weaken the utility of the pairwise kSZ measurement, especially when a prior on $\sigma_z$ is provided from the broader DES analysis program.

\subsection{Results}
\label{sec:results}

\begin{figure*}
\begin{center}
\includegraphics[width=1.0\textwidth]{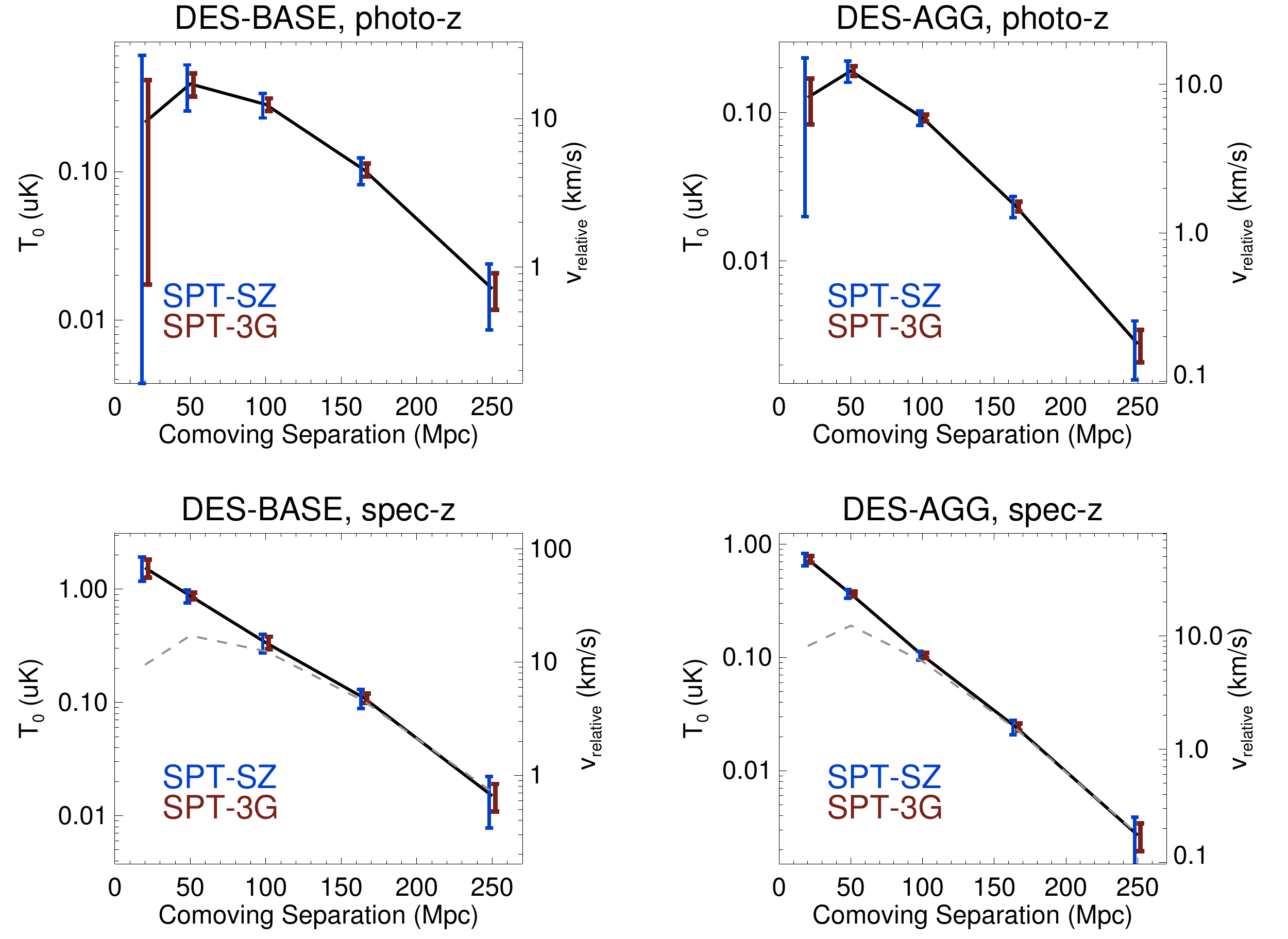}
\end{center}
\caption{The simulated pairwise kSZ signal for the \des~(\textit{left panels}) and \desagg~(\textit{right panels}) samples, and for photometric (\textit{top panels}) and spectroscopic (\textit{bottom panels}) redshifts.  The black curve shows the mean signal, while the error bars show the projected uncertainties using data from \sptsz~(\textit{blue, left offset}) and \sptg~(\textit{red, right offset}).  The \textit{dashed, gray} curve in the bottom panels shows the mean signal using photometric redhifts, to highlight the loss of signal at \dcom$\lesssim$ 50 Mpc and the preservation of signal at \dcom$\gtrsim$ 50 Mpc.  The errors include contributions from instrumental noise, a background of point sources, the CMB, the kSZ signal itself, and residual tSZ signal, and are mildly correlated between bins of cluster separation.  The second vertical axis shows the corresponding velocity per cluster, assuming optical depths of $\tau=0.0025$ for \des~and $\tau=0.0017$ for \desagg, which are typical of the clusters used in each sample.\vskip 30pt}
\label{fig:snr}
\end{figure*}

We use the simulations described in the previous sections to determine the detection significance of the pairwise kSZ as a function of pair-averaged redshift.  For each redshift bin we compute $\chi^2 = d^{\rm{T}} C d$, where $d$ is the mean kSZ signal as a function of pair separation in this redshift bin and $C$ is the covariance matrix described above, and calculate SNR~$\equiv\sqrt{\sum_i \chi^2_i}$, where the sum is over the redshift bins.  We summarize the results in Table~\ref{tab:snr} and Figure~\ref{fig:snr}.

We find that existing data from the \sptsz~survey will, when combined with either of the DES samples considered here, result in a strong detection of the pairwise kSZ signal.  The \des~(\desagg) sample yields a 8.2$\sigma$ (12.9$\sigma$) detection.  For both samples, the measurement is largely limited by instrumental noise in the \sptsz~data.


As expected, we find that data from a future, third-generation survey, \sptg, would result in even more precise measurements of the pairwise kSZ signal.  The \des~(\desagg) sample yields a 18$\sigma$ (30$\sigma$) detection.  In contrast to \sptsz, the \sptg~constraints are limited by kSZ and tSZ noise, with roughly equal contributions from each.

We considered using a combination of 95 and 150 GHz \sptg~data to eliminate the tSZ signal and its associated variance, but found this was not worth the effective increase in instrumental noise; the resulting detection significance on the \desagg~sample was 18$\sigma$, weaker than the 30$\sigma$ obtained using the original analysis.

Finally, we explored how these constraints would change with spectroscopic followup of the DES cluster sample, such as would be possible with DESpec \citep{abdalla12}.  We assumed a spectroscopic redshift error of $\sigma_z=0.002$ per cluster and found that \sptsz~(\sptg) would yield a 12.5$\sigma$ (26$\sigma$) detection on the \des~sample and a 26$\sigma$ (43$\sigma$) detection on the \desagg~sample.

\begin{table*}
\begin{center}
\begin{threeparttable}
\caption[]{Sensitivity to pairwise kSZ signal}
\begin{tabular}{ | c | c | c | c | c |}
\hline \hline
 & \des~(photo-$z$) & \desagg~(photo-$z$) & \des~(spec-$z$) & \desagg~(spec-$z$)\\
\hline
\sptsz & 8.2$\sigma$~(12\%) & 12.9$\sigma$~(7.8\%) & 12.5$\sigma$~(8.0\%) & 21$\sigma$~(4.8\%) \\
\hline
\sptg & 18$\sigma$~(5.5\%) & 30$\sigma$~(3.3\%) & 26$\sigma$~(3.9\%) & 43$\sigma$~(2.3\%) \\
 \hline \hline
\end{tabular}
\label{tab:snr}
\begin{tablenotes}
\item The projected detection significance of the pairwise kSZ measurement for different combinations of CMB data and cluster samples.  \vskip 10pt 
\end{tablenotes}
\end{threeparttable}
\end{center}
\end{table*}


\subsection{Implications for Testing Gravity}
\label{sec:gravity}

The pairwise kSZ signal directly measures the average pairwise ionized momentum of a given cluster sample.  The results of the previous section demonstrate that precision measurements of this signal should be available in the coming years, and these have the power to yield detailed information on the gas content of the DES clusters.

To go beyond gas physics --- to constrain cosmology or gravity --- we must convert to average pairwise \textit{velocity}.  This requires an independent measurement of the average ionized gas profile, say from tSZ or X-ray observations.  
Additionally, if there is a well motivated connection between the total cluster mass and
the ionized gas mass, then measurements of the mean mass of the sample
through weak lensing or angular clustering can also constrain the conversion from
kSZ signal to velocity.
For the remainder of this section we assume that the uncertainty in the gas as constrained by independent data is subdominant to the uncertainty on the pairwise kSZ measurement, an assumption that seems plausible at least for the $\sim$10\% SPT-SZ kSZ measurement.

The mean pairwise velocity of halos $\< v_h \>(r)$ as measured through
the kSZ effect as function of
separation $r$ is in linear perturbation theory given by \citep{schmidt10}
\ba
\< v_h \>(r) = 2\, \bar b\, \xi_{\delta v}(r).
\ea
Here, $\bar b$ is the kSZ signal-weighted mean of $(b-1)$ of halos 
in the given 
mass bin, $b$ is the linear Eulerian bias, and $\xi_{\delta v}(r)$ is the
linear cross-correlation between density and velocity.    
If the kSZ signal is simply proportional to
gas mass which is proportional to halo mass, $\bar b$ reduces to the
mass-weighted average of $b-1$.  We assume that there is no velocity 
bias of halos, which holds on the large scales of interest.  

The velocity-matter cross correlation is given by
\be
\xi_{\delta v}(r,z) = - \frac{a H f}{2\pi^2} \int dk\: k P(k,z) j_1(kr),
\ee
where $f = d\ln D/d\ln a$ is the growth rate,
$P(k,z)$ is the matter power spectrum at redshift $z$, and $j_l$
denote spherical Bessel functions.  

\begin{figure*}
\begin{center}
\includegraphics[width=0.65\textwidth]{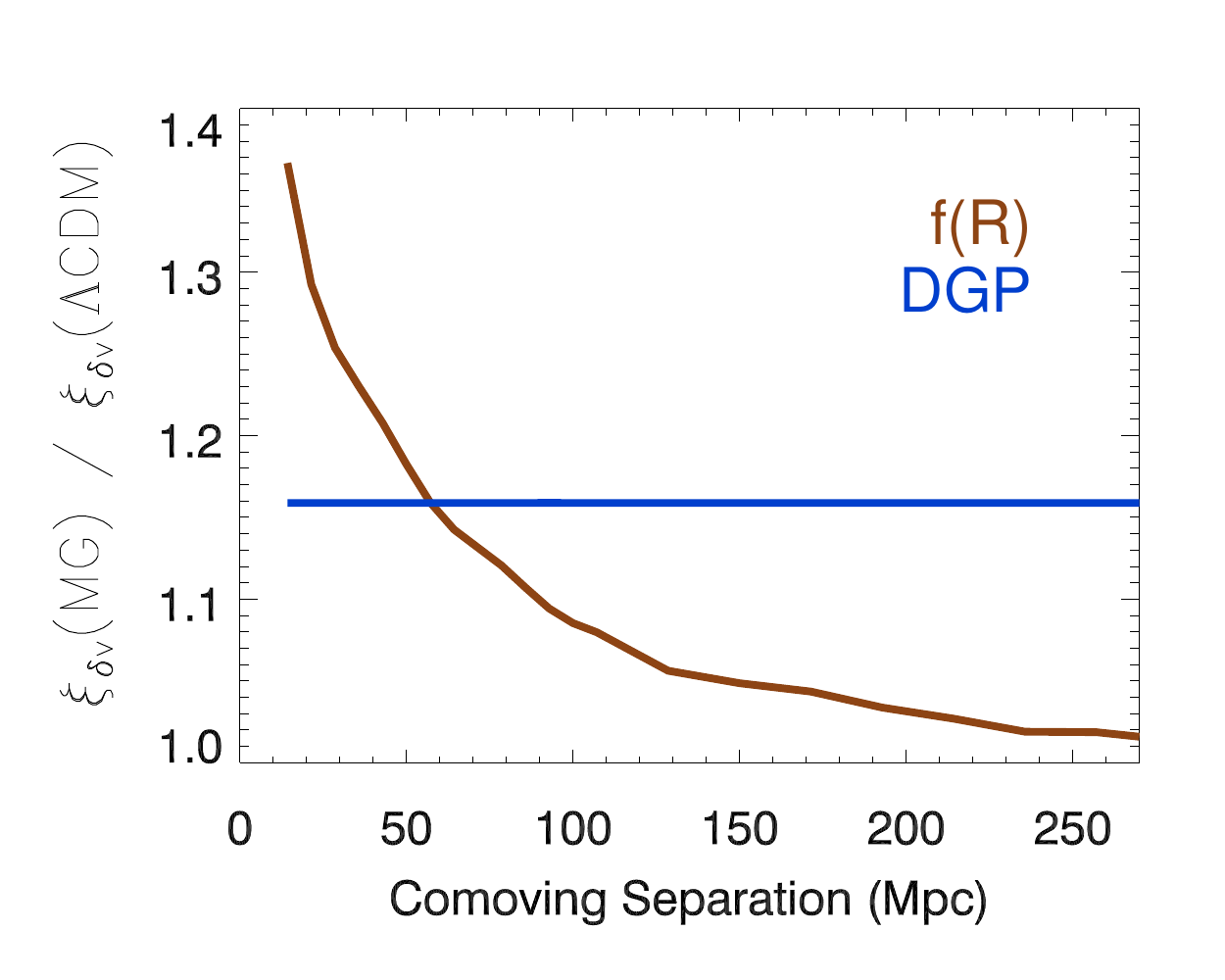}
\end{center}
\caption{The change in the density-velocity correlation function $\xi_{\delta v}$ in two models of modified gravity.  The $f(R)$ model uses $|f_{R0}|=10^{-4}$, the maximum field value currently allowed.  The DGP model uses $r_c = 4000$ Mpc.  \vskip 30pt}
\label{fig:MG}
\end{figure*}

Consider a modified gravity (MG) model that leads to a scale-independent
modification of growth on linear scales (one example is the
Dvali-Gabadadze-Porrati (DGP) model \citep{dvali2000} and its generalizations).  
The effect of such a modification of gravity on the velocity-matter
cross-correlation at fixed redshift is given by
\be
\frac{\xi_{\delta v,\rm MG}(r,z)}{\xi_{\delta v, \Lambda\rm  CDM}(r,z)} =
\frac{[f \sigma^2_8(z)]_{\rm MG}}{[f \sigma^2_8(z)]_{\Lambda\rm CDM}},
\label{eq:xiratio}
\ee
where $\sigma^2_8(z)$ is the power spectrum normalization at redshift $z$.  
On linear scales, the mean pairwise velocity as measured through
kSZ thus constrains $\bar b f\, \sigma^2_8$.  Independent knowledge of $\bar b$
can then be used to turn the kSZ measurement into a constraint
on $f \sigma_8^2$.  When including constraints on the expansion history
and primordial amplitude of fluctuations, this can then be turned into
a constraint on modified gravity.  Specifically, for the normal-branch
DGP model with \LCDM expansion history considered in \cite{schmidt09}
which is parametrized by the cross-over scale $r_c$, 
a measurement of $\xi_{\delta v}(r)$ with an overall SNR of 10 (20) --- as should be possible using the DES cluster sample in conjunction with existing data from \sptsz~(future data from \sptg) --- would 
yield a 95\% confidence level lower limit of $r_c \gtrsim$ 3000 (7000) Mpc, assuming that the expansion history, halo bias, and kSZ-velocity relation 
are known perfectly (Fig.~\ref{fig:MG}).
This is an interesting constraint, since a modified gravity explanation
for the accelerated expansion of the Universe suggests 
 $r_c \sim 3000 h^{-1}$~Mpc as a natural value in these models. 
 
We note that uncertainty in the cluster gas should affect only the global normalization of the kSZ signal considered here, not the \textit{shape} of the kSZ signal as a function of separation.  If this claim is supported by simulations, it could be used to test the scale-dependence of gravity on sufficiently large scales, independently of gas physics.  
A well-studied class of modified gravity models which includes the 
$f(R)$ model \citep{carrolletal} invokes a massive additional degree of freedom.  The 
modifications to the gravitational force and velocities are generically
scale-dependent and suppressed on scales larger than the Compton length of the field,
typically 30~Mpc or less (see Fig.~\ref{fig:MG}).
For this reason, only cluster samples with spectroscopic redshifts are useful for constraining $f(R)$.  Our most optimistic scenario, in which \sptg~observes the spectroscopically followed-up \desagg~cluster sample, results in a 15$\sigma$ (6.7\%) detection of pairwise kSZ signal in the the 10-30 Mpc separation bin.  For comparison, the maximum currently-allowed value of the field amplitude in the \cite{HuSawicki} $f(R)$ model is $|f_{R0}|=10^{-4}$ \citep{schmidtetal09,lombriseretal10}, and results in a $\sim$25\% increase in $\xi_{\delta v}$ over this range.  The \sptg+\desagg(spec-z) scenario would therefore allow a 4$\sigma$ test of this hypothesis.


\section{Discussion}
\label{sec:discussion}

We have estimated the sensitivity of mm-wave data from SPT to the pairwise kSZ signal of clusters from the DES.  We find that existing SPT data will allow for an 8-13$\sigma$ detection of the kSZ signal.  If the kSZ-velocity relation is known perfectly, this corresponds to an 8-12\% constraint on $f\sigma_8^2$,  which is comparable to the current best constraint on growth from redshift space distortions (RSD), an 8.2\% constraint on $f\sigma_8$ from BOSS \citep{reid12}.  
 
The difference is that the bulk of the RSD sensitivity arises from quasi-linear scales \citep{reid11}, whereas the SPT+DES kSZ signal is most sensitive to significantly larger, more easily modeled, linear scales.  This is due to the DES photometric redshifts, and spectroscopic followup would boost the kSZ SNR by a factor of 1.5.  A third-generation SPT survey would improve any of these constraints by a factor of $\sim$2.1, and is limited largely by noise from the kSZ and tSZ signals.  These measurements have the potential to provide interesting constraints on theories of modified gravity.

\begin{acknowledgments}
We thank Tom Crawford, Lloyd Knox, Gil Holder, Eduardo Rozo, and Christian Reichardt for conversations regarding this work, and the University of Texas at Austin --- where this work was initiated --- for its hospitality.  RK acknowledges support from NASA Hubble Fellowship grant HF-51275 and NSF grant ANT-0959620, and FS from the NASA Einstein Fellowship.  This research used resources from the RCC at the University of Chicago and NERSC, which is supported by the U.S. DOE under Contract DE-AC02-05CH11231 \end{acknowledgments}


\end{document}